# Superfluidity of a perfect quantum crystal

V. A. Golovko

Moscow State Evening Metallurgical Institute

Lefortovsky Val 26, Moscow 111250, Russia

E-mail: mgvmi-mail@mtu-net.ru

Abstract

In recent years, experimental data were published which point to the possibility of the existence of superfluidity in solid helium. To investigate this phenomenon theoretically we employ a hierarchy of equations for reduced density matrices which describes a quantum system that is in thermodynamic equilibrium below the Bose-Einstein condensation point, the hierarchy being obtained earlier by the author. It is shown that the hierarchy admits solutions relevant to a perfect crystal (immobile) in which there is a frictionless flow of atoms, which testifies to the possibility of superfluidity in ideal solids. The solutions are studied with the help of the bifurcation method and some their peculiarities are found out. Various physical aspects of the problem, among them experimental ones, are discussed as well.

**1. Introduction**

The question as to the possibility of the existence of superfluidity in a crystal solid has come under scrutiny long ago [1-3]. However, rather firm experimental evidence for this possibility in solid $^4$He was obtained only in recent years (for a review see [4] where the experimental as well as theoretical aspects of the question are discussed).

In the present paper, the superfluid state of a crystalline solid is investigated with the help of the approach in equilibrium quantum statistical mechanics proposed in [5] (see also [6] where a systematic exposition of the approach is presented and various results achieved with its use are discussed). The approach is based upon a hierarchy of equations for reduced density matrices obtained in the same paper [5], the hierarchy going over, in the classical limit, into the well-known equilibrium Bogolyubov-Born-Green-Kirkwood-Yvon (BBGKY) hierarchy. A characteristic feature of the approach is construction of thermodynamics compatible with the hierarchy without use made of the Gibbs method. In Ref. [7] that serves as starting point for the present investigation, a modification of the hierarchy of [5] was obtained to take account of the Bose-Einstein condensation in fluids.

It should be emphasized that the superfluidity of a perfect crystal is implied in the present paper whereas it is common practice to ascribe superfluidity in a solid to disruptions of the ideal crystalline order (vacancies, interstitials) [4,8–10]. It is also worth adding that Balibar and Caupin [8] conclude, upon analyzing various papers along the latter lines, that the question as to whether a perfect crystal can be superfluid is still a matter of controversy. The results of the present paper show that the perfect crystal can be superfluid, and the approach used provides a means for studying this phenomenon. This can help in establishing the actual mechanism of superfluidity in solids. At the same time it should be remarked that different mechanisms are not excluded in different experimental situations.

In this paper, theoretical treatment of the problem will first be carried out. In the concluding section of the paper, various physical aspects of the problem, among them experimental ones, will be discussed as well.

**2. Basic equations**

We consider a system of $N$ spinless bosons ($^4$He atoms, for example) enclosed in a volume $V$. The particles of mass $m$ interact via a two-body potential $K(|\mathbf{r}_j - \mathbf{r}_k|)$. Let us first present a synopsis of the approach developed in Ref. [5][1]. According to [5] we introduce $s$-particle reduced density matrices $R_s(\mathbf{x}_s, \mathbf{x}'_s)$ with $s = 1, 2, 3, \ldots$ where $\mathbf{x}_s$ denotes a set of coordinates $\mathbf{r}_1, \mathbf{r}_2, \ldots, \mathbf{r}_s$. The approach proceeds on the hierarchy of equations for $R_s$ which follows from the quantum mechanical Liouville equation for $R_N$ [12,13]. We look for a stationary solution to the hierarchy that should correspond to thermodynamic equilibrium in the form

$$R_s(\mathbf{x}_s, \mathbf{x}'_s) = \sum_\nu n_s\left(\varepsilon_\nu^{(s)}\right) \psi_\nu(\mathbf{x}_s) \psi_\nu^*(\mathbf{x}'_s), \qquad (2.1)$$

the functions $\psi_\nu(\mathbf{x}_s)$ satisfying the equations

$$\mathsf{H}^{(s)} \psi_\nu(\mathbf{x}_s) = \varepsilon_\nu^{(s)} \psi_\nu(\mathbf{x}_s) \quad \text{with} \quad \mathsf{H}^{(s)} = -\frac{\hbar^2}{2m} \sum_{j=1}^{s} \nabla_j^2 + U_s(\mathbf{x}_s), \qquad (2.2)$$

in which $U_s(\mathbf{x}_s)$ are functions assumed to be symmetric in the coordinates. They play the role of effective potentials. We imply symmetric solutions of Eq. (2.2) as we consider bosons.

---

[1] In this reference spinless fermions are treated as well, incorporation of the spin into the approach is considered in [11]; see also [6].

If Eq. (2.1) is placed in the above-mentioned hierarchy and Eq. (2.2) is used, in the limit as $\mathbf{x}'_s \to \mathbf{x}_s$ we arrive at

$$\rho_s(\mathbf{x}_s) \nabla_1 U_s(\mathbf{x}_s) = \rho_s(\mathbf{x}_s) \nabla_1 \sum_{j=2}^{s} K\left(|\mathbf{r}_1 - \mathbf{r}_j|\right) + \int \rho_{s+1}(\mathbf{x}_{s+1}) \nabla_1 K\left(|\mathbf{r}_1 - \mathbf{r}_{s+1}|\right) d\mathbf{r}_{s+1}, \qquad (2.3)$$

where $\rho_s(\mathbf{x}_s) \equiv R_s(\mathbf{x}_s, \mathbf{x}_s)$ are diagonal elements of the density matrices:

$$\rho_s(\mathbf{x}_s) = \sum_\nu n_s\left(\varepsilon_\nu^{(s)}\right) |\psi_\nu(\mathbf{x}_s)|^2. \qquad (2.4)$$

It is to be added that the following normalization for the density matrices is implied

$$\int_V \rho_1(\mathbf{r}) d\mathbf{r} = N. \qquad (2.5)$$

Considerations that permit one to suggest that the form of $R_s(\mathbf{x}_s, \mathbf{x}'_s)$ as given by (2.1) should correspond to a state of thermodynamic equilibrium are expounded in [5] (see also [6]). In any case, from the mathematical point of view the closed hierarchy of stationary equations which is given by Eqs. (2.2)–(2.4) and which contains only the diagonal elements of the density matrices is a direct consequence of the quantum mechanical Liouville equation derived without using any approximation. The noteworthy fact is that the functions $n_s(z)$ with $z = \varepsilon_\nu^{(s)}$ are arbitrary in the above stationary hierarchy. Once the hierarchy is solved, the full density matrices can be reconstructed by (2.1).

It may be noted that the reduced density matrices can always be put in a form analogous with (2.1) by linear transformation of the functions $\psi_\nu(\mathbf{x}_s)$. The resulting representation of the reduced density matrices is often used in different studies [14,15]. Eq. (2.1) implies another thing, namely, at thermodynamic equilibrium the diagonalization of the coefficients in the series occurs for all $s$ simultaneously if use is made of the functions $\psi_\nu(\mathbf{x}_s)$ that satisfy the Schrödinger type stationary equations of (2.2). Besides, we supposed that the coefficients $n_s$ in (2.1) depend upon $\nu$ only via $\varepsilon_\nu^{(s)}$. To put it otherwise, the functions $n_s(z)$ do not depend on $\nu$ (see also [6]).

It is not obligatory, however, that any solution to the hierarchy obtained will really represent reduced density matrices[2]. In compliance with their definition [5], the reduced density matrices must be interrelated by

$$(N - s + 1) R_{s-1}(\mathbf{x}_{s-1}, \mathbf{x}'_{s-1}) = \int_V R_s(\mathbf{x}_{s-1}, \mathbf{r}_s, \mathbf{x}'_{s-1}, \mathbf{r}_s) d\mathbf{r}_s, \quad s = 2, 3, \ldots . \tag{2.6}$$

There is also another condition on these matrices. If a system consists of two mutually noninteracting subsystems $A$ and $B$, and the wavefunction is of the form

$$\Psi(\mathbf{x}_N) = \Psi_A \Psi_B, \tag{2.7}$$

all reduced densities matrices should break up into two factors each of which corresponds to $A$ or $B$, which also follows from the definition of the reduced density matrices.

In Ref. [5] where uniform media are treated, the case is considered in which the functions $n_s(z)$ are smooth. In this case, instead of (2.1), another representation for $R_s(\mathbf{x}_s, \mathbf{x}'_s)$ can be obtained which is more convenient for practical use (Eq. (2.2) too acquires another form). The conditions embodied by (2.6) and (2.7) enable one to find the functions $n_s(z)$:

$$n_s(z) = A s! \rho^{s-1} \left( \frac{2\pi\hbar^2}{m\tau} \right)^{3(s-1)/2} e^{-z/\tau}, \tag{2.8}$$

where $A$ and $\tau$ are parameters independent of $z$, and $\rho = N/V$. The parameter $A$ is fixed by the normalization condition of (2.5). As to $\tau$, this parameter should depend on the temperature $\theta$ and the density $\rho$. To find $\tau$ it is necessary to construct thermodynamics compatible with the hierarchy obtained. This is done in Ref. [5] in which an equation that specifies the function $\tau(\theta,\rho)$ is deduced as well.

Examples of solution of the hierarchy equations considered in [5] demonstrate that in the case of Bose systems the function $\tau(\theta,\rho)$ becomes negative at sufficiently low temperatures whereas it should be positive according to (2.8). It is clear that this inconsistency is due to a manifestation of Bose-Einstein condensation, and the approach used is to be modified to take this phenomenon into account. This is done in Ref. [7] for uniform media, and henceforth we shall reason in parallel with [7] (see also [6]) up to the point where the fact that we imply a crystal should be taken into explicit account.

It is not necessary for the functions $n_s(\varepsilon_\nu^{(s)})$ in (2.1) to be smooth. Let us suppose now that the functions have an outlier at a certain $\nu = \nu_0$ depending on the number $s$ (in actual fact, $\nu$ is a set of quantum numbers different for different $s$ [6]), which amounts to saying that $n_s(\varepsilon_{\nu_0}^{(s)}) = \Delta n_s + n_s(z_0)$ where $n_s(z_0)$ is the limit of $n_s(z)$ as $z \to z_0 = \varepsilon_{\nu_0}^{(s)}$. The part of the system described by $\Delta n_s$ will be called the condensate according to the commonly accepted terminology. In this case, Eq. (2.1) can be recast as

$$R_s(\mathbf{x}_s, \mathbf{x}'_s) = R_s^{(c)}(\mathbf{x}_s, \mathbf{x}'_s) + R_s^{(n)}(\mathbf{x}_s, \mathbf{x}'_s), \qquad (2.9)$$

where $R_s^{(c)}(\mathbf{x}_s, \mathbf{x}'_s)$ is relevant to the condensate while $R_s^{(n)}(\mathbf{x}_s, \mathbf{x}'_s)$ to the normal fraction. The normal part $R_s^{(n)}(\mathbf{x}_s, \mathbf{x}'_s)$ is represented by the same equation as (2.1) with smoothed functions $n_s(z)$. The condensate part $R_s^{(c)}(\mathbf{x}_s, \mathbf{x}'_s)$ comes from the term in (2.1) that contains $\Delta n_s$. We incorporate $\sqrt{\Delta n_s}$ into $\psi_{\nu_0}(\mathbf{x}_s)$ and denote the function thus obtained as $\varphi_s(\mathbf{x}_s)$, so that

$$R_s^{(c)}(\mathbf{x}_s, \mathbf{x}'_s) = \varphi_s(\mathbf{x}_s)\,\varphi_s^*(\mathbf{x}'_s). \qquad (2.10)$$

---

[2] In studies on density matrices, such a situation is known as the *N*-representability problem [16].

The normalization of the functions $\varphi_s(\mathbf{x}_s)$ is unknown for the moment because we do not know $\Delta n_s$. The functions satisfy the equation that follows immediately from (2.2):

$$\frac{\hbar^2}{2m}\sum_{j=1}^{s}\nabla_j^2\,\varphi_s(\mathbf{x}_s) + \left[\varepsilon_{(s)} - U_s(\mathbf{x}_s)\right]\varphi_s(\mathbf{x}_s) = 0, \qquad (2.11)$$

in which $\varepsilon_{(s)}$ is written for $\varepsilon_{\nu_0}^{(s)}$.

As a result we see that in the presence of the condensate the reduced density matrices break up into a sum of the condensate and normal parts as given by Eq. (2.9). This separation is exact in the representation employed in the present approach inasmuch as here again no approximation was used. It is worthy of remark that Eq. (2.3) for the effective potentials $U_s(\mathbf{x}_s)$ contains the complete diagonal elements $\rho_s(\mathbf{x}_s) = \rho^{(c)}(\mathbf{x}_s) + \rho^{(n)}(\mathbf{x}_s)$. In addition to the unknown functions $n_s(z)$, the hierarchy in the present case contains other unknown quantities, namely, the constants $\varepsilon_{(s)}$ in Eq. (2.11), the normalization of $\varphi_s(\mathbf{x}_s)$ and the boundary conditions for the differential equation of (2.11). All of these can be found from the conditions given by (2.6) and (2.7).

First of all we remark that the interrelation (2.6) is linear and therefore we can require it to be satisfied by $R_s^{(n)}(\mathbf{x}_s, \mathbf{x}'_s)$ and $R_s^{(c)}(\mathbf{x}_s, \mathbf{x}'_s)$ separately. Consideration of the case of two mutually noninteracting subsystems when Eq. (2.7) holds, together with the interrelation (2.6) applied to $R_s^{(n)}(\mathbf{x}_s, \mathbf{x}'_s)$, will again lead to (2.8). This is proven strictly in the full version of Ref. [7]. It remains now to consider the interrelation (2.6) as applied to the condensate part $R_s^{(c)}(\mathbf{x}_s, \mathbf{x}'_s)$. Substituting (2.10) into (2.6) yields

$$\varphi_{s-1}(\mathbf{x}_{s-1})\,\varphi_{s-1}^*(\mathbf{x}'_{s-1}) = \frac{1}{N}\int_V \varphi_s(\mathbf{x}_{s-1},\mathbf{r}_s)\,\varphi_s^*(\mathbf{x}'_{s-1},\mathbf{r}_s)\mathrm{d}\mathbf{r}_s, \qquad (2.12)$$

where account has been taken of the fact that $s \ll N$ in the case under study.

The following analysis necessitates some properties of the effective potentials $U_s(\mathbf{x}_s)$. Eq. (2.3) defines them up to arbitrary constants $C_s$. On a base of Eq. (2.3), in Appendix C of Ref. [5] it is strictly shown that $U_s(\mathbf{x}_s) \to U_{s-1}(\mathbf{x}_{s-1}) + U_1(\mathbf{r}_s) + C_s$ as $|\mathbf{r}_s| \to \infty$. At the same time, when deriving an equation that leads to (2.8), a remark following Eq. (4.9) of [5] was made according to which in the case of a crystal one should set $C_s = -\bar{U}_1$ where $\bar{U}_1$ is the constant part of $U_1(\mathbf{r})$. As a result, if $|\mathbf{r}_s| \to \infty$ we have

$$U_s(\mathbf{x}_s) \to U_{s-1}(\mathbf{x}_{s-1}) + U_1(\mathbf{r}_s) - \bar{U}_1. \tag{2.13}$$

We turn now to Eq. (2.11) with $s = 1$:

$$\frac{\hbar^2}{2m}\nabla^2 \varphi_1(\mathbf{r}) + \left[\varepsilon_{(1)} - U_1(\mathbf{r})\right]\varphi_1(\mathbf{r}) = 0. \tag{2.14}$$

Up to this point we have nowhere taken into explicit account the fact that we are interested in solutions relevant to a crystal. The uniform case where $U_1(\mathbf{r})$ = constant was studied in [7]. In the case of the crystal, the potential $U_1(\mathbf{r})$ must be periodic. Eq. (2.14) is of the form of a Schrödinger equation whereas the character of solutions of the Schrödinger equation for a periodic potential is well known [17]:

$$\varphi_1(\mathbf{r}) = \sqrt{\rho_c}\, e^{i\frac{\mathbf{p}_0}{\hbar}\mathbf{r}}\, u_1(\mathbf{r}), \tag{2.15}$$

where $u_1(\mathbf{r})$ is a periodic function with the same periods as the potential $U_1(\mathbf{r})$. Usually, in the exponent one writes $i\mathbf{k}\mathbf{r}$; however it is more convenient for us to write $\mathbf{p}_0/\hbar$ instead of $\mathbf{k}$. Eq. (2.15) contains a yet unknown normalizing factor $\sqrt{\rho_c}$ introduced with the condition that the function $u_1(\mathbf{r})$ obeys the normalization

$$\int_V |u_1(\mathbf{r})|^2\, d\mathbf{r} = V. \tag{2.16}$$

It is worth remarking that $u_1 = 1$ in the uniform case [7], which conforms to (2.16). For use later, we also write down Eq. (2.11) at $s = 2$:

$$\frac{\hbar^2}{2m}\left[\nabla_1^2 \varphi_2(\mathbf{r}_1,\mathbf{r}_2) + \nabla_2^2 \varphi_2(\mathbf{r}_1,\mathbf{r}_2)\right] + \left[\varepsilon_{(2)} - U_2(\mathbf{r}_1,\mathbf{r}_2)\right]\varphi_2(\mathbf{r}_1,\mathbf{r}_2) = 0. \tag{2.17}$$

Let us try and satisfy the condition of (2.12) at the least possible $s = 2$:

$$\varphi_1(\mathbf{r}_1)\, \varphi_1^*(\mathbf{r}_1') = \frac{1}{N}\int_V \varphi_2(\mathbf{r}_1,\mathbf{r}_2)\, \varphi_2^*(\mathbf{r}_1',\mathbf{r}_2)d\mathbf{r}_2. \tag{2.18}$$

At given $\mathbf{r}_1$ and $\mathbf{r}'_1$, the main contribution to the last integral in the thermodynamic limit as $V \to \infty$ results from the regions where $|\mathbf{r}_1-\mathbf{r}_2|$ and $|\mathbf{r}'_1-\mathbf{r}_2|$ are large. If $|\mathbf{r}_1-\mathbf{r}_2| \to \infty$, Eq. (2.13) yields

$$U_2(\mathbf{r}_1,\mathbf{r}_2) \to U_1(\mathbf{r}_1) + U_1(\mathbf{r}_2) - \bar{U}_1. \tag{2.19}$$

If (2.19) is placed in (2.17), the variables in the resulting equation will be separated and the solution of the equation, in view of Eq. (2.14), will be of the form

$$\varphi_2^{(\infty)}(\mathbf{r}_1,\mathbf{r}_2) = B_2\varphi_1(\mathbf{r}_1)\varphi_1(\mathbf{r}_2), \qquad (2.20)$$

where $B_2$ is a constant and the superscript ($\infty$) underlines that this is the limiting value of $\varphi_2(\mathbf{r}_1,\mathbf{r}_2)$. Simultaneously, we find that

$$\varepsilon_{(2)} = 2\varepsilon_{(1)} - \bar{U}_1. \qquad (2.21)$$

According to the foregoing, it is Eq. (2.20) that can be inserted into (2.18) in case the limit $V \to \infty$ is implied. Then the integral is calculated at once on account of (2.15) and (2.16) to yield

$$B_2 = \sqrt{\frac{\rho_0}{\rho_c}}, \qquad (2.22)$$

where $\rho_0 = N/V$ (without loss of generality the quantity $B_2$ can be taken to be real). Hence we have found $\varepsilon_{(2)}$ according to (2.21) and established that the solution of Eq. (2.17) should be subject to the limiting condition of (2.20) that, together with (2.22), determines the normalization of $\varphi_2(\mathbf{r}_1,\mathbf{r}_2)$.

Arbitrary $s$ can be treated in a like manner. The main contribution to the integral in (2.12) in the thermodynamic limit is given by the regions where $|\mathbf{r}_s| \to \infty$. In this limit, we have (2.13). If this is inserted into (2.11) and account is taken of (2.14) and of Eq. (2.11) written for $\varphi_{s-1}(\mathbf{x}_{s-1})$, by analogy with (2.20) and (2.21) we shall obtain

$$\varphi_s^{(\infty)}(\mathbf{x}_s) = B_s\varphi_{s-1}(\mathbf{x}_{s-1})\varphi_1(\mathbf{r}_s), \quad \varepsilon_{(s)} = \varepsilon_{(s-1)} + \varepsilon_{(1)} - \bar{U}_1. \qquad (2.23)$$

Putting this limiting function into (2.12) yields, analogously to (2.22),

$$B_s = \sqrt{\frac{\rho_0}{\rho_c}}. \qquad (2.24)$$

Thus, we have completely constructed the hierarchy of equations for the reduced density matrices which describes the crystal below the Bose-Einstein condensation point that can consequently be observed in the crystalline state as well. The normal part of the reduced density matrices is described by the equations for $R_s^{(n)}(\mathbf{x}_s,\mathbf{x}'_s)$ of [7] and [5] whereas their condensate part is determined by Eqs. (2.10)-(2.11) whose solutions should be subject to the first condition of (2.23) with $B_s$ of (2.24). The constants $\varepsilon_{(s)}$ are to be found from the second equation of (2.23) which enables one to express them via $\varepsilon_{(1)}$ that is a function of $\mathbf{p}_0$ specified by Eq. (2.14) The condensate and normal parts are interconnected by the effective potentials $U_s(\mathbf{x}_s)$ found from Eq. (2.3). At the same time, this last equation links the $s$th and ($s + 1$)th members of the hierarchy. The hierarchy obtained contains the arbitrary constants $\rho_c$ and $\mathbf{p}_0$ as well as two constants ($A$ and $\tau$) in Eq. (2.8).

It should be emphasized once more that no approximation was used when deriving the hierarchy. The hierarchy obtained is valid in the thermodynamic limit ($N \to \infty$ and $V \to \infty$ with $N/V$ = constant), which is seen from the above derivation and from Ref. [5] whose results should be employed for the normal part of the density matrices. The hierarchy is derived on assuming that $s \ll N$.

Let us make three remarks as to the hierarchy obtained. The presence of $\rho_c$ in the denominator of (2.22) and (2.24) should not cause difficulties in case the limit as $\rho_c \to 0$ is considered. The fact is that Eqs. (2.20) and (2.23) contain $\varphi_1(\mathbf{r}_s)$ as a factor while the function $\varphi_1(\mathbf{r}_s)$ itself has a factor $\sqrt{\rho_c}$ in view of (2.15). As a result, the factor $\sqrt{\rho_c}$ disappears in fact from Eq. (2.23) that relates $\varphi_s(\mathbf{x}_s)$ and $\varphi_{s-1}(\mathbf{x}_{s-1})$, so that all $\varphi_s(\mathbf{x}_s)$'s will have $\sqrt{\rho_c}$ as a factor just as $\varphi_1(\mathbf{r}_s)$.

The second remark consists in the following. Equations of the type (2.14) with a periodic potential $U_1(\mathbf{r})$ are used in studies of movement of particles, e. g. of an electron, in a periodic field [17], as was mentioned above. In this case there exists an infinite set of values of $\varepsilon_{(1)}$ that form energy bands if the vector $\mathbf{p}_0$ (in our notation) changes. In our case, one should take the band that corresponds to the minimum of an appropriate thermodynamic potential while all subsequent $\varepsilon_{(s)}$'s will be obtained uniquely by (2.23). The situation here is analogous with that which occurs in the theory of an ordinary crystal. At a given interaction potential between particles, a great diversity of crystalline lattices can exist. Realized is the lattice that corresponds to a minimum of the relevant thermodynamic potential (see, for example, [18]).

The third remarks concerns the equations for $R_s^{(n)}(\mathbf{x}_s, \mathbf{x}'_s)$. Inasmuch as the condensate and the normal fraction are linked by the common potentials $U_s(\mathbf{x}_s)$, those equations should also have solutions corresponding to a periodic density once $U_1(\mathbf{r})$ is periodic. Such solutions do exist and are considered in Ref. [19]. They have properties characteristic of an ordinary crystal.

It is of interest to find the singlet density matrix. To this end, we substitute (2.10) with $s = 1$ into (2.9) and take account of (2.15), which gives

$$R_1(\mathbf{r},\mathbf{r}') = \rho_c \, e^{i\frac{\mathbf{p}_0}{\hbar}(\mathbf{r}-\mathbf{r}')} u_1(\mathbf{r}) u^*_1(\mathbf{r}') + R_1^{(n)}(\mathbf{r},\mathbf{r}'), \tag{2.25}$$

where $R_1^{(n)}(\mathbf{r},\mathbf{r}')$ describes the normal fraction. The first term in (2.25) exhibits off-diagonal long range order (ODLRO) since it does not vanish in the limit as $|\mathbf{r} - \mathbf{r}'| \to \infty$. It is worthy of remark that ODLRO is characteristic of the phenomenon of superfluidity [20]. The second term in (2.25) does not display ODLRO as in an ordinary crystal, which can be proven rigorously. In the uniform case, the fact that this term tends to zero as $|\mathbf{r} - \mathbf{r}'| \to \infty$ is seen from Eq. (2.32) of Ref. [21] in which superfluidity of a Fermi liquid is considered.

In order to elucidate the physical sense of the results obtained let us calculate the momentum of the system that is given by Eq. (2.13) of [7]:

$$\mathbf{P} = -i\hbar \int \left[ \nabla R_1(\mathbf{r},\mathbf{r}') \right]_{\mathbf{r}'=\mathbf{r}} d\mathbf{r}. \qquad (2.26)$$

Upon substituting the first term of (2.25) herein (the second term should not contribute to the macroscopic momentum as in an ordinary crystal), we get

$$\mathbf{P} = \rho_c \mathbf{p}_0 V - i\hbar \rho_c \int_V u_1^*(\mathbf{r}) \nabla u_1(\mathbf{r}) d\mathbf{r}, \qquad (2.27)$$

where (2.16) has been taken into account. Hence the immobile crystal (its density $\rho_1(\mathbf{r})$ is time-independent) has a momentum that can be relevant only to a movement of the condensate. As long as the number of particles in the condensate is $N_c = \rho_c V$, upon dividing $\mathbf{P}$ by $mN_c$ we obtain the mean velocity of the directional movement of the condensate particles although this does not signify that all condensate particles move simultaneously. At the same time, one cannot, of course, distinguish between the moving particles and the others owing to the principle of indistinguishability of identical particles. It should be emphasized that the flow of particles in the crystal is not accompanied by any dissipation for the system is in thermodynamic equilibrium. This result is completely analogous to the one obtained for uniform media in Ref. [7] (in the case of the uniform media the last term in (2.27) is absent). We shall return to the physical aspect of the result in the concluding section of the present paper.

It should be added also that it may turn out that $\mathbf{p}_0 = 0$. Then we shall have a condensate phase ($\rho_c \neq 0$) without superfluidity (if $\mathbf{p}_0 = 0$, the second term in (2.27) vanishes as well, see Sec. 4). If $\mathbf{p}_0 \neq 0$, the condensate phase will be superfluid. Thus, formation of the condensate phase does not necessarily leads to superfluidity just as in the case of uniform media [7]. It should be stressed that this conclusion implies a state of thermodynamic equilibrium. Even if the thermodynamically equilibrium state is nonsuperfluid ($\mathbf{p}_0 = 0$), the equations obtained above always admit solutions with $\mathbf{p}_0 \neq 0$. The solutions can correspond to superfluid metastable states. An example of this is provided by the results of Ref. [22] in which the present approach is applied for studying systems of spinless bosons bound by forces of gravity alone under conditions of Bose-Einstein condensation (at absolute zero of temperature). The ground state (the state with the least possible energy) corresponds to an immobile structure and is not superfluid. The rotating structure is superfluid but its energy is higher. At the same time, the lifetime of this excited state can be infinite owing to the conservation of angular momentum as noted in [22]. We shall return to this question in the concluding section of the paper.

The next step in the employed approach of [5] consists in construction of thermodynamics, which enables one simultaneously to obtain equations for determination of the quantities $\rho_c$ and $\mathbf{p}_0$ as well as of the two constants that characterize $R_s^{(n)}(\mathbf{x}_s, \mathbf{x}'_s)$. In the uniform case, the thermodynamics is built up in [7]. As to a crystal, the construction of thermodynamics is essentially complicated by an involved form of the pair correlation function in the crystal [23] while the function enters into expressions for thermodynamic quantities. Up to the present, even in the case of an ordinary crystal it was possible to construct thermodynamics in the framework of the approach only with use made of simplifying assumptions as to the form of the pair correlation function, and the quantum case [19] is noticeably more complicated than the classical [24]. The presence of the condensate adds complexity to the construction of thermodynamics. Inasmuch as in the present paper we are interested first of all in the possibility in principle concerning the existence of superfluidity in a crystal, in what follows we shall restrict ourselves to the case of zero temperature ($\theta = 0$) where thermodynamics is not required and the equilibrium state of the system is determined by the condition that its energy is a minimum.

## 3. Periodic solutions

Having in mind the case of zero temperature (see the end of the preceding section) and trying to simplify, wherever possible, the problem under study, we shall presume that all particles pertain to the condensate, that is to say, $\rho_c = \rho_0 \equiv N/V$. It should be remarked that this occurs in the case of the Bose liquid considered in [7], although in an analogous case of the Fermi liquid this does not happen [21]. If $\rho_c = \rho_0$, the complete reduced density matrices coincide with $R_s^{(c)}(\mathbf{x}_s, \mathbf{x}'_s)$ while the spatial number density of particles is (henceforth we shall omit the subscript 1 of $\rho_1$, $u_1$ and $U_1$)

$$\rho(\mathbf{r}) = \rho_0 |u(\mathbf{r})|^2, \tag{3.1}$$

where (2.15) has been taken into account (note that (3.1) agrees with the normalization condition of (2.5) owing to (2.16)). If (2.15) is placed in (2.14), there results the equation for $u(\mathbf{r})$

$$\frac{\hbar^2}{2m}\nabla^2 u(\mathbf{r}) + \frac{i\hbar \mathbf{p}_0}{m}\nabla u(\mathbf{r}) + \left[\varepsilon_{(1)} - \frac{\mathbf{p}_0^2}{2m} - U(\mathbf{r})\right] u(\mathbf{r}) = 0. \tag{3.2}$$

The effective potential $U(\mathbf{r})$ is determined by Eq. (2.3) at $s = 1$ which is of the form

$$\rho(\mathbf{r}) \nabla U(\mathbf{r}) = \int \rho_2(\mathbf{r}, \mathbf{r}') \nabla K(|\mathbf{r} - \mathbf{r}'|) d\mathbf{r}'. \tag{3.3}$$

We introduce also the pair correlation function $g(\mathbf{r}, \mathbf{r}')$ in line with the customary relation

$$\rho_2(\mathbf{r}, \mathbf{r}') = \rho(\mathbf{r}) \rho(\mathbf{r}') g(\mathbf{r}, \mathbf{r}'). \tag{3.4}$$

The function $\rho_2(\mathbf{r},\mathbf{r}')$ and thereby $g(\mathbf{r},\mathbf{r}')$ should be found from the subsequent hierarchy equations at $s = 2$ that contain $\rho_3(\mathbf{r}_1,\mathbf{r}_2,\mathbf{r}_3)$ as well. Putting aside the discussion of the problem as to how to close the hierarchy, the problem well known from the studies of the classical BBGKY hierarchy, for the moment we shall assume the function $g(\mathbf{r},\mathbf{r}')$ to be given. Almost in all studies on statistical theory of crystals, one introduces a simplifying assumption that $g(\mathbf{r},\mathbf{r}')$ depends only on $|\mathbf{r} - \mathbf{r}'|$ as in fluids. This assumption should not essentially affect emerging results because in the case of a crystal the leading role for $\rho_2(\mathbf{r},\mathbf{r}')$ of (3.4) is played by the periodic density $\rho(\mathbf{r})$ while $g(\mathbf{r},\mathbf{r}')$ plays a secondary role [18]. If $g(\mathbf{r},\mathbf{r}') = g(|\mathbf{r} - \mathbf{r}'|)$, Eq. (3.3) is readily integrated to give

$$U(\mathbf{r}) = \int K_g(|\mathbf{r}-\mathbf{r}'|)\rho(\mathbf{r}')d\mathbf{r}', \qquad (3.5)$$

where

$$K_g(r) = \int_\infty^r \frac{dK(r')}{dr'} g(r')dr'. \qquad (3.6)$$

Here $r = |\mathbf{r}|$ and the limits of integration in (3.6) are chosen such that $K_g(\infty) = 0$ (only in this case the integral in (3.5) converges). It will be remarked that an arbitrary constant can de added to the right side of (3.5), which defines another choice for the origin of the energy scale. In case $g(r)$ is taken to be given, one has two equations (3.2) and (3.5) that allow one to find the density in the crystal by (3.1).

We look for periodic solutions of Eq. (3.2) in terms of a Fourier series:

$$u(\mathbf{r}) = \sum_{l,m,n=-\infty}^{\infty} c_{lmn} e^{i\mathbf{A}\mathbf{r}}, \qquad (3.7)$$

where $\mathbf{A} = l\mathbf{a}_1 + m\mathbf{a}_2 + n\mathbf{a}_3$ with the basic reciprocal-lattice vectors $\mathbf{a}_1$, $\mathbf{a}_2$ and $\mathbf{a}_3$ [24]. Eq. (2.16) entails the following condition on the coefficients $c_{lmn}$ in the limit as $V \to \infty$:

$$\sum_{l,m,n} |c_{lmn}|^2 = 1. \qquad (3.8)$$

The crystal density of (3.1) is represented by the series

$$\rho(\mathbf{r}) = \sum_{l,m,n} a_{lmn} e^{i\mathbf{A}\mathbf{r}}, \qquad a_{lmn} = \rho_0 \sum_{l',m',n'} c_{l'm'n'} c^*_{l'-l,m'-m,n'-n}. \qquad (3.9)$$

The noteworthy fact is that $a_{000} = \rho_0$ in view of (3.8). Substituting (3.9) into (3.5) yields

$$U(\mathbf{r}) = \sum_{l,m,n} a_{lmn} \sigma(A) e^{i\mathbf{A}\mathbf{r}}, \qquad (3.10)$$

where

$$\sigma(k) = \int K_g(|\mathbf{r}|)\,e^{i\mathbf{k}\mathbf{r}}\,d\mathbf{r} = \frac{4\pi}{k}\int_0^\infty r K_g(r)\sin kr\,dr = \frac{4\pi}{k^3}\int_0^\infty \frac{dK}{dr} g(r)(kr\cos kr - \sin kr)\,dr, \quad (3.11)$$

and $A = |\mathbf{A}|$. The last expression was obtained upon inserting $K_g(r)$ of (3.6) and carrying out one of the integrations.

If all these series are put into (3.2), we shall arrive at the following set of equations that contain the coefficients $c_{lmn}$ alone:

$$\left[\frac{(\hbar \mathbf{A} + \mathbf{p}_0)^2}{2m} + \rho_0 \sigma_0 - \varepsilon_{(1)}\right]c_{lmn} + \rho_0 \sum_{l',m',n'}{}' \sigma(A') c_{l-l',m-m',n-n'} \sum_{l''m''n''} c_{l''m''n''}\, c^*_{l''-l',m''-m',n''-n'} = 0. \quad (3.12)$$

Here the primed summation denotes the omission of the term with $l' = m' = n' = 0$ because this term has been separated out and gives the summand $\rho_0\sigma_0$ in the square brackets with use made of (3.8) and with the notation $\sigma_0 = \sigma(0)$.

Let us find the expression for the momentum of the crystal in the present case. Substituting (3.7) into (2.27) and retaining only terms that increase with the volume $V$ yields

$$\mathbf{P} = N\left(\mathbf{p}_0 + \hbar \sum_{l,m,n} \mathbf{A}|c_{lmn}|^2\right). \quad (3.13)$$

It follows herefrom that in general the momentum $\mathbf{P}$ of the crystal is not aligned with the vector $\mathbf{p}_0$.

We turn now to the energy of the crystal. In the approach used, the energy of the system is specified by the formula [5]

$$E = -\frac{\hbar^2}{2m}\int \left[\nabla^2 R_1(\mathbf{r},\mathbf{r}')\right]_{\mathbf{r}'=\mathbf{r}} d\mathbf{r} + \frac{1}{2}\int K(|\mathbf{r}_1-\mathbf{r}_2|)\rho_2(\mathbf{r}_1,\mathbf{r}_2)\,d\mathbf{r}_1\,d\mathbf{r}_2. \quad (3.14)$$

We substitute the first term of (2.25) here, transform the resulting expression with the help of (3.2), and place the above Fourier series in the expression obtained. As a result, we have (cf. the derivation of Eq. (4.3) in [24])

$$E = \varepsilon_{(1)}N - V\sum_{l,m,n}|a_{lmn}|^2 \sigma_\varepsilon(A), \quad \sigma_\varepsilon(k) = \int\left[K_g(|\mathbf{r}|) - \tfrac{1}{2}K(|\mathbf{r}|)g(|\mathbf{r}|)\right]e^{i\mathbf{k}\mathbf{r}}\,d\mathbf{r}. \quad (3.15)$$

In an analogous way, one can calculate the stress tensor in the present case starting from the general formula (3.5) of [7].

It is desirable to estimate the pair correlation function $g(r)$ that figures in the above formulae. In different statistical theories of crystals, especially in the ones based upon the density functional (see the reviews [25,26]), beginning with the pioneering work on the statistical theory of a (classical) crystal [27], for $g(r)$ is usually taken the pair correlation function of the corresponding liquid. We shall proceed along these lines. If $\rho_c = \rho_0$, according to [7] the condensate part of the pair correlation function of the liquid is $g(\mathbf{r}) = |u_2(\mathbf{r})|^2$ while the function $u_2(\mathbf{r})$ is to be found from Eq. (2.11) of [7]. This last equation contains the potential $U_2(\mathbf{r})$ that depends upon the triplet correlation function by virtue of Eq. (5.33) of [5], and therefore we have to resort to an approximation. We shall start from the quantum extension of the hypernetted-chain approximation proposed in [5]. The relevant equation (5.36) of [5] contains the quantity $\tau$ whose role goes over to $\tilde{\tau} = (1 - \rho_c/\rho_0)\tau$ below the Bose-Einstein condensation point [7,21]. Because of this, in our case where $\rho_c = \rho_0$ and thereby $\tilde{\tau} = 0$, one should set $\tau = 0$ in Eq. (5.36) of [5], which gives immediately that $U_2(r) = K(r)$ (it may be noted in passing that one obtains the same result from the classical hypernetted-chain equation (5.35) of [5] at $\theta = 0$; the equality $U_2(r) = K(r)$ results also at any temperature when the triplet correlations are neglected [5]). After the replacement $U_2(r) = K(r)$ in the above-mentioned equation (2.11) of [7], in the spherically symmetric case we arrive at

$$\frac{d^2 u_2}{dr^2} + \frac{2}{r}\frac{du_2}{dr} - \frac{m}{\hbar^2} K(r) u_2(r) = 0. \tag{3.16}$$

This equation must be solved with the condition that $u_2(r) \to 1$ as $r \to \infty$.

For the atomic interaction potential, we take the Lennard-Jones potential

$$K(r) = \varepsilon\left[\left(\frac{r_m}{r}\right)^{12} - 2\left(\frac{r_m}{r}\right)^6\right], \tag{3.17}$$

in which $r_m$ corresponds to the minimum of $K(r)$ and $K(r_m) = -\varepsilon$, and which passes through zero at $r = r_0 = r_m/2^{1/6}$. The potential has a strong singularity as $r \to 0$, which considerably complicates numerical solution of Eq. (3.16). For this reason we shall proceed as follows. We take some $r_1 < r_0$ and set $K(r) = \infty$ if $r < r_1$, that is to say, we shall use a hard-sphere potential for $r < r_1$. When $r > r_1$, we shall solve Eq. (3.16) with the potential (3.17) and with the condition $u_2(r_1) = 0$. In numerical calculation, we took $r_1 = 0.85 r_0$ where the potential (3.17) is sufficiently large: $K(r_1) \approx 18\varepsilon$. At lesser values of $r_1$, the above-mentioned singularity of the potential (3.17) begins to manifest itself markedly.

If (3.17) is put into Eq. (3.16) and the equation is reformulated in terms of dimensionless quantities, in front of the dimensionless potential there appears the factor

$$\zeta = \frac{m\varepsilon r_m^2}{\hbar^2}. \tag{3.18}$$

For helium-4, $\varepsilon = 10.2$ K, $r_m = 2.86$ Å [28] and thereby $\zeta = 6.9$. Eq. (3.16) with this $\zeta$ was solved numerically with the help of the well-known Runge-Kutta method. We took $u_2(r_1) = 0$ and by the trial-and-error method we looked for a value of the derivative $u'_2(r_1)$ such that $u_2(r) \to 1$ as $r \to \infty$. The results of the calculation are presented in figure 1 where $\tilde{r} = r/r_m$.

Now we can compute the function $\sigma(k)$ of (3.11) that is the sole prescribed function in the set of equations of (3.12). The form of the function is shown in figure 2 where $\tilde{\sigma} = \sigma/(\varepsilon r_m^3)$ and $\tilde{k} = kr_m$. The noteworthy fact is that this form of $\sigma(k)$ is fully analogous with the one depicted in figure 1 of Ref. [23] in the classical case for zero temperature (that figure contains an error: in actual fact it is the quantity $\sigma/(\varepsilon r_m^3) = \tilde{\sigma}$ that is laid off on the vertical axis). The values characteristic of a crystal are $\tilde{k} \sim 2\pi$ [23]. They lie between points $A$ and $B$ in figure 2 where $\sigma(k) < 0$, which is necessary for the crystal to exist as will be shown in the next section.

## 4. Bifurcation method

One of the methods employed for solving nonlinear equations describing a crystal is the bifurcation method. Although the bifurcation point is of little physical significance, the method enables one to find out the conditions for periodic solutions of the equations to exist and to investigate some their properties [24]. The method consists in searching for periodic solutions characterizing the crystal that bifurcate off the uniform one relevant to the corresponding liquid.

In our case, the set of (3.12) has a solution $c_{000} = 1$ with the remaining $c_{lmn} = 0$, which corresponds to a liquid (this can, more simply, be seen from Eqs. (3.2) and (3.5) that admit a solution $u(\mathbf{r}) = 1$, $\rho(\mathbf{r}) = \text{constant}$, $U(\mathbf{r}) = \text{constant}$). We now look for a solution where some $c_{lmn}$ connected by the symmetry are small upon assuming the remaining $c_{lmn}$ to be of higher order of magnitude except for $c_{000} \approx 1$ by (3.8). If we take Eq. (3.12) at $l = m = n = 0$ ($\mathbf{A} = 0$) and discard all small terms (such will be all terms under the summation sign), in a zeroth approximation we shall get

$$\varepsilon_{(1)} = \frac{\mathbf{p}_0^2}{2m} + \rho_0 \sigma_0 . \tag{4.1}$$

To obviate any confusion we note that this $\varepsilon_{(1)}$ differs by the last term from $\varepsilon_{(1)}$ written for the uniform case in Eq. (2.8) of [7], which is due to a different choice for the origin of the energy scale (see the remark concerning (3.5)).

In the remaining equations of (3.12), we now retain only terms of the first order of magnitude (recall that $c_{000} \approx 1$, Eq. (4.1) is also used) with the result that

$$\left(\frac{\hbar^2}{2m}A^2 + \frac{\hbar}{m}\mathbf{p}_0\mathbf{A}\right)c_{lmn} + \rho_0\sigma(A)\left(c^*_{-l,-m,-n} + c_{lmn}\right) = 0. \quad (4.2)$$

Upon changing the sign of *l*, *m*, *n* we take the complex conjugate:

$$\left(\frac{\hbar^2}{2m}A^2 - \frac{\hbar}{m}\mathbf{p}_0\mathbf{A}\right)c^*_{-l,-m,-n} + \rho_0\sigma(A)\left(c_{lmn} + c^*_{-l,-m,-n}\right) = 0. \quad (4.3)$$

The set of these two homogeneous equations for $c_{lmn}$ and $c^*_{-l,-m,-n}$ has a nontrivial solution only if the determinant of the set is equal to zero. Before calculating the determinant it is worthwhile to establish what $c_{lmn}$'s are connected by the symmetry. For the sake of simplicity, we shall imply cubic lattices (SC, FCC or BCC of the simplest symmetry) although solid $^4$He has a hexagonal close-packed (HCP) lattice at low pressures. However the difference in energy between the FCC and HCP structures are small; besides, $^4$He can have a BCC or FCC lattice at some pressures and temperatures [28]. In addition, the Lennard-Jones potential used of (3.17) describes the interaction between helium atoms rather roughly in order to lead to the HCP lattice unequivocally. The aforementioned cubic lattices have a different number of basic coefficients $c_{lmn}$ that correspond to the vectors $\mathbf{a}_i$ obtainable from the basic reciprocal-lattice vectors $\mathbf{a}_1$, $\mathbf{a}_2$ and $\mathbf{a}_3$ by the relevant symmetry transformations, all these $\mathbf{a}_i$'s having the same magnitude $|\mathbf{a}_i| = a$ (see, e.g., [24,19]). We substitute this *a* into the set of Eqs. (4.2)–(4.3) instead of *A*, put $\mathbf{A} = \mathbf{a}_i$ and equate the determinant of the set to zero, so that

$$\frac{\hbar^2}{2m}a^2 - \frac{2p_0^2\cos^2\xi_i}{m} + 2\rho_0\sigma(a) = 0, \quad (4.4)$$

where $\xi_i$ is the angle between the vectors $\mathbf{p}_0$ and $\mathbf{a}_i$. This equation is the condition of appearance of nonzero $c_{lmn}$'s, that is to say, the condition under which a periodic solution branches off from the uniform one (the bifurcation condition).

Inasmuch as the lattice period is inversely proportional to *a*, the average density of the crystal $\rho_0$ is proportional to $a^3$, which amounts to saying that $\rho_0 = a^3/\eta$ where $\eta = \eta'\pi^3$ with $\eta' = 8$, $6\sqrt{3}$, $8\sqrt{2}$ for the SC, FCC, BCC lattices respectively. We now rewrite Eq. (4.4) as

$$\sigma(a) = -\frac{\hbar^2\eta}{2ma} + \frac{p_0^2\eta\cos^2\xi_i}{ma^3}. \quad (4.5)$$

This equation is conveniently solved graphically. We consider first the case $p_0 = 0$. In this case, from (4.5) it follows immediately that for the crystal to exist it is necessary that $\sigma(a) < 0$. We are interested in solutions between points *A* and *B* in figure 2 (see the end of the preceding section). The right-hand side of (4.5) at $p_0 = 0$ is plotted by the broken curve in this figure when the solutions required occur. It should not be surprising that two solutions rather than one exist necessarily in this instance. To clarify the situation, let us resort to the classical case where the bifurcation condition has a form similar to (4.4) at $p_0 = 0$ if the quantity $\hbar^2 a^2/m$ is replaced by a quantity proportional to the temperature (see Eq. (6.3) of [24]), and thereby has two solutions as well. Account must be taken of the fact that the different values of $a$ correspond to different pressures; besides, only stable solutions are of importance (the solutions that provide a minimum for the relevant thermodynamic potential). The solutions of equations for the classical crystal obtained in [18] not only in the vicinity of the bifurcation point but also in the entire region where the solutions exist show that at a given pressure and given temperature there is a unique stable solution of a specified symmetry.

It remains now to be seen whether there are relevant solutions in the helium case as long as the required solutions will be lacking in case the broken curve of figure 2 passes below the minimum of $\sigma(k)$ between points $A$ and $B$. To this end, we recast Eq. (4.5) at $p_0 = 0$ in the dimensionless form:

$$\tilde{\sigma}(a) = -\frac{\eta}{4\zeta a r_m}, \qquad (4.6)$$

where the dimensionless parameter $\zeta$ of (3.18) is used. According to the foregoing, $\eta \sim 10\pi^3$ for the cubic lattices, $ar_m \sim 2\pi$ between points $A$ and $B$ in figure 2, and $\zeta = 6.9$ for helium. Therefore the right side of Eq. (4.6) between these points is of the order of $-2$ whereas the minimum value of $\tilde{\sigma}(\tilde{k})$ is equal to $-4.5$ there. Consequently, the situation in helium corresponds to that presented in figure 2, and the required periodic solutions do exist. At the same time, it is to be remarked that the broken curve of figure 2 for helium passes not too far from the minimum of $\tilde{\sigma}(\tilde{k})$. If a more realistic interatomic potential is used instead of (3.17) and if the function $u_2(r)$ that should depend on the pressure is calculated with use made of more sophisticated approximations, it may turn out that at low pressures the broken curve of figure 2 will pass below the curve for $\tilde{\sigma}(\tilde{k})$. This will explain the fact that at low pressures helium remains liquid down to zero temperature. However, to explain the fact convincingly it needs to investigate Eq. (3.12) in case $c_{lmn}$'s are not small. As to other rare gases, the parameter $\zeta$ of (3.18) for them is substantially greater because their values of $m$, $\varepsilon$ and $r_m$ far exceed those of helium. As a result, according to (4.6) the broken curve of figure 2 for them will pass much closer to the abscissa and far from the minimum of $\sigma(k)$, so that the influence of the pressure should not be so crucial. For this reason, these rare gases solidify prior to attaining zero temperature at any pressure.

We turn now to the case $p_0 \neq 0$. If $p_0$ is small in (4.5), the broken curve of figure 2 will rise but slightly, and the periodic solutions will exist as before. However, the magnitude of $a = |\mathbf{a}_i|$ will now depend upon the angle $\xi_i$, and the lattice will cease to be cubic. Such deformation of the lattice and the consequent change of the crystal symmetry at a nonzero $p_0$ are quite natural. If in a perfect crystal there exists a polar vector (the vector $\mathbf{P}$ of (2.27) in our case), the crystal symmetry must be such that the vector is admitted. Only 10 pyroelectric crystal classes allow of such a vector [29], and the cubic and HCP crystals do not figure among them. Therefore, in the presence of a superflow in these last crystals, their symmetry should change. The question as to whether this conclusion remains valid if the superflow is due to crystal imperfections requires special investigation since the symmetry considerations are not applicable, strictly speaking, in this case.

When $p_0$ is sufficiently large, the broken curve in figure 2 will rise substantially and the required solutions with $\cos\xi_i \neq 0$ may disappear at all. However, for some directions of the vector $\mathbf{p}_0$ with respect to the vectors $\mathbf{a}_i$, when $\cos\xi_i = 0$, the periodic solutions will remain although the periodicity will not be three-dimensional now. Insofar as the last term in (4.5) at $p_0\cos\xi_i \neq 0$ tends to $+\infty$ as $a \to 0$ ($\propto 1/a^3$), there exist solutions with $p_0 \neq 0$ and small $a$, that is, long-period structures, even if the solutions in the vicinity of the interval $AB$ of figure 2 are lacking. On the other hand, if the solutions in the vicinity of the interval $AB$ are nonexistent at $p_0 = 0$, they may appear at $p_0 \neq 0$ when the broken curve will rise sufficiently in conformity with (4.5). Thus, there is a theoretical possibility of the existence of periodic structures only with the simultaneous presence of a superflow.

In the formula for the momentum of the crystal of (3.13), in the present approximation it needs to sum over the above-specified vectors $\mathbf{a}_i$ alone. We express $c^*_{lmn}$ in terms of $c_{-l,-m,-n}$ with the help of (4.2) upon changing the sign of $l$, $m$, $n$ and substitute into the sum of (3.13). With regard to the fact that the vectors $\mathbf{a}_i$ and $-\mathbf{a}_i$, to which the coefficients $c_{lmn}$ and $c_{-l,-m,-n}$ correspond, are of the same magnitude because (4.4) contains $\cos^2\xi_i$, some terms in the sum obtained will cancel out. As a result, we shall have

$$\mathbf{P} = N\mathbf{p}_0 + \frac{2\hbar^2 V}{m} \sum_i {}' \frac{\mathbf{a}_i}{\sigma(a_i)} \mathbf{p}_0 \mathbf{a}_i c_{lmn} c_{-l,-m,-n}, \qquad (4.7)$$

where the primed summation signifies that only one of the two vectors $\mathbf{a}_i$ or $-\mathbf{a}_i$ should be taken into account, no matter which. Eq. (4.7) demonstrates explicitly that the vector $\mathbf{P}$ need not be directed along $\mathbf{p}_0$ although its magnitude is proportional to $p_0$.

Subsequent calculation in the bifurcation method should be carried out separately for each type of the lattice. The calculations can be performed by analogy with an ordinary crystal, the classical [24] or quantum [19]. We shall not carry out the calculations as long as their results are not required for the aims of the present paper.

Of importance is the question as to whether the solutions with $p_0 \neq 0$ can correspond to a minimum of energy, that is to say, whether they can represent the ground state of the system. If (4.1) is substituted into (3.15), we shall see that the energy increases with increasing $p_0$ at small $|a_{lmn}|$. This, however, does not necessarily mean that the same will occur at $|a_{lmn}|$ relevant to a real crystal. By way of example we can point out the classical crystal where the crystalline state becomes energetically advantageous only at sufficiently large $|a_{lmn}|$ [18]. It is quite possible that advantageous will be superflows only with particular orientations of the vector $\mathbf{p}_0$ with respect to the crystal axes. To answer these questions, other approaches to solving the set of equations of (3.12) should be searched for besides the bifurcation method.

## 5. Concluding remarks

The present paper shows that a perfect quantum crystal can possess superfluidity. Let us discuss how this phenomenon can be understood from the physical point of view. The superfluidity of a crystal may be conceived as a peculiar kind of collective tunnelling of the crystal particles when the points corresponding to the density maximums remain immobile to form a regular crystalline lattice though deformed by the superflow. At low temperatures, macroscopic bodies tend to have a crystalline order; on the other hand, helium tends to become superfluid. The results of the present paper demonstrate that these two tendencies do not contradict each other.

It should be observed that in the present study a perfect crystal is implied in which the number of atoms $N$ is equal to the number of lattice sites $N_s$. At the same time, when discussing superfluidity of solid helium the possibility of an incommensurate crystal in which $N \neq N_s$ is considered as well [4,8]. In the approach used in the present study it is not at all imperative that $N = N_s$. In Sec. 4 of Ref. [18] where a BCC lattice (space group $O_h^9$) is considered, it was found that, on lowering the temperature, the ordinary BCC lattice can go over into a BCC lattice in which the number of lattice sites exceeds that of particles. This last lattice is, however, metastable because its free energy is greater than that of the ordinary BCC lattice. Hence, the present approach provides a means of studying crystals in which $N \neq N_s$ as well.

The approach of this paper enables one to treat both the superfluidity of a liquid and the one of a crystal in perfect analogy. It all depends on whether we look for uniform or periodic solutions of Eq. (2.14). From the viewpoint of the approach, the superflow can exist irrespective of whether the atoms of the substance are located at random (the liquid) or in a perfect order (the crystal). Because of this, one can, for the crystal, utilize the analogies and arguments from the concluding section of Ref. [7] that will not be repeated here.

Let us take a brief look at the following question alone. We considered an unbounded crystal in the present paper. In finite crystalline specimens, the superflows must close upon themselves somehow as in liquids [7]. It is quite possible that the specimen will break up into cells with closed superflows. Perhaps, this explains the negative result obtained when trying to detect mass transport in solid helium [30]. At the same time, it should be emphasized once again that the superfluid solid does not flow like a fluid. It remains immobile and solid while the superflow exists in the interior of the solid (analogously with the flow of electrons that move in an immobile conductor that carries a current). When trying to detect the mass transport in a superfluid crystal experimentally, the challenge is to bring out the superflow through the surface of the specimen. This may be done, for example, if the supersolid is in contact with a superfluid on two sides. A superflow from the superfluid can penetrate the supersolid in which some of the closed superflows can open to escape through the second side. In this connection let us cite Ref. [31] in which experimental observation of mass transport in solid $^4$He that is in contact with superfluid $^4$He was reported (see, however, [32,33]).

To prove more convincingly that a superfluid state of solid helium is observed in some experiment or other, it is desirable to investigate the crystal lattice of helium by crystallographic methods. If, for example, the mass transport observed by Ray and Hallock [31] is due to liquid channels in the specimen [32], the crystalline structure of solid helium should remain unchanged. If, however, the superflow traverses the helium crystal, the crystal symmetry should change according to Sec. 4 of the present paper.

It is not excluded also that the state with the least possible energy, that is, the ground state will be the one with no superflow ($p_0 = 0$) whereas the state with a superflow ($p_0 \neq 0$) will be excited (the same may occur in a liquid as well [7]). Under certain external conditions, this excited state may be metastable and may exist for an appreciable length of time. The state may even become ground. In rotatory movement, for example, the kinetic energy of a body is $E_k = I\omega^2/2$ where $I$ is the moment of inertia of the body and $\omega$ is its angular velocity. If a superflow develops, the moment of inertia of the "normal" part of the body rotating with the angular velocity $\omega$ decreases, and the velocity of the "supurfluid" part decreases as well if the superflow is opposite in direction to the rotation. As a result, the total energy of the rotating body may become less than at $p_0 = 0$. The situation here is analogous to that with a ferromagnet in which formation of magnetic domains is energetically unfavourable from the structural point of view, nevertheless the domains do form because this leads to a decrease in the magnetic energy of the specimen. It is worth remarking that superfluidity in solid $^4$He was observed for the first time with confidence in torsional oscillator experiments [4] for which the above considerations are valid. It should be added that the abovementioned observation of mass transport in solid $^4$He that is in contact with superfluid $^4$He [31] may be explained also in the case where the ground state of the crystal is not superfluid. The superflow injected from the superfluid into the crystal can cause the crystal to pass into an excited state that is superfluid.

Of interest is to discuss the possibility of superfluidity in solid $^3$He whose atoms are fermions as distinct from the bosons considered in this paper. The superfluidity of a fermionic liquid in the framework of the approach proposed in [5] was considered in [21]. Although the treatment of the problem proves to be more involved than in the case of spinless bosons, by and large the ideas of Ref. [7] that were exploited in the present paper remain in force. In particular, Eq. (2.21) of [21] will coincide completely with Eq. (2.14) of the present paper if account is taken of the potential $U_1(\mathbf{r})$ that can be set (and was set) equal to zero in the case of the liquid. Other equations will also coincide if $U_1(\mathbf{r})$ is allowed for. Therefore, there exists the possibility in principle as to the existence of superfluidity in solid $^3$He as well.

It is not inconceivable that the superfluid state may occur in other crystals as well, especially in rare gas solids other than solid helium. There is no general formula for the Bose-Einstein condensation temperature, below which superfluidity can exist, even in the case of liquids, let alone the solids. As is well known, rather a good estimate for the λ-transition temperature in liquid helium is given by the formula for the Bose-Einstein condensation temperature in an ideal gas (3.1 K instead of the observed temperature 2.2 K):

$$T_c = \frac{3.31\hbar^2}{k_B m}\left(\frac{N}{V}\right)^{2/3}, \qquad (5.1)$$

in which $k_B$ is the Boltzmann constant, $m$ is the particle mass, and spinless bosons are implied. In the case of $^{20}$Ne, for example, at zero temperature and pressure the nearest-neighbour distance in the lattice is $b = 3.16$ Å [28], which enables one to calculate the number density $N/V = 1/v_0$ where $v_0$ is the volume per atom with $v_0 = b^3/\sqrt{2}$ for the FCC lattice. Now, Eq. (5.1) yields $T_c = 1.0$ K for neon. This temperature is not too low. It is to be observed that the estimate as given by (5.1) may become more exact at high pressures when the role of the attractive part of the interaction potential diminishes [28], so that the behaviour of the rare gas solids should more closely resemble the behaviour of solid helium.

123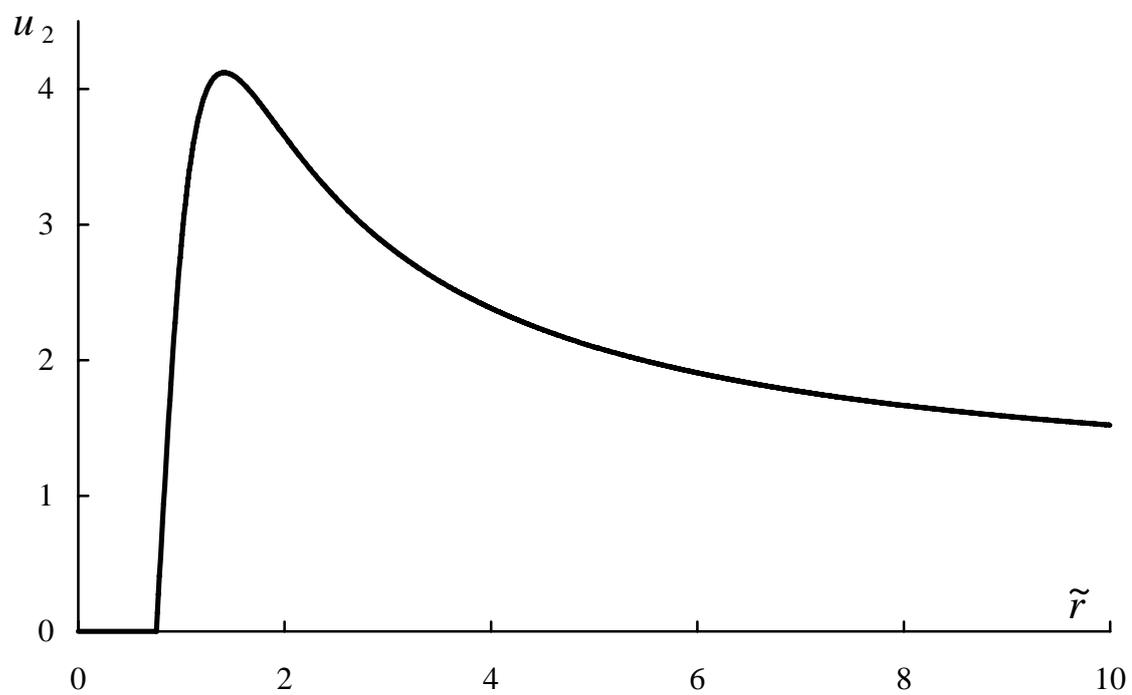

**Figure 1**. Function $u_2(r)$

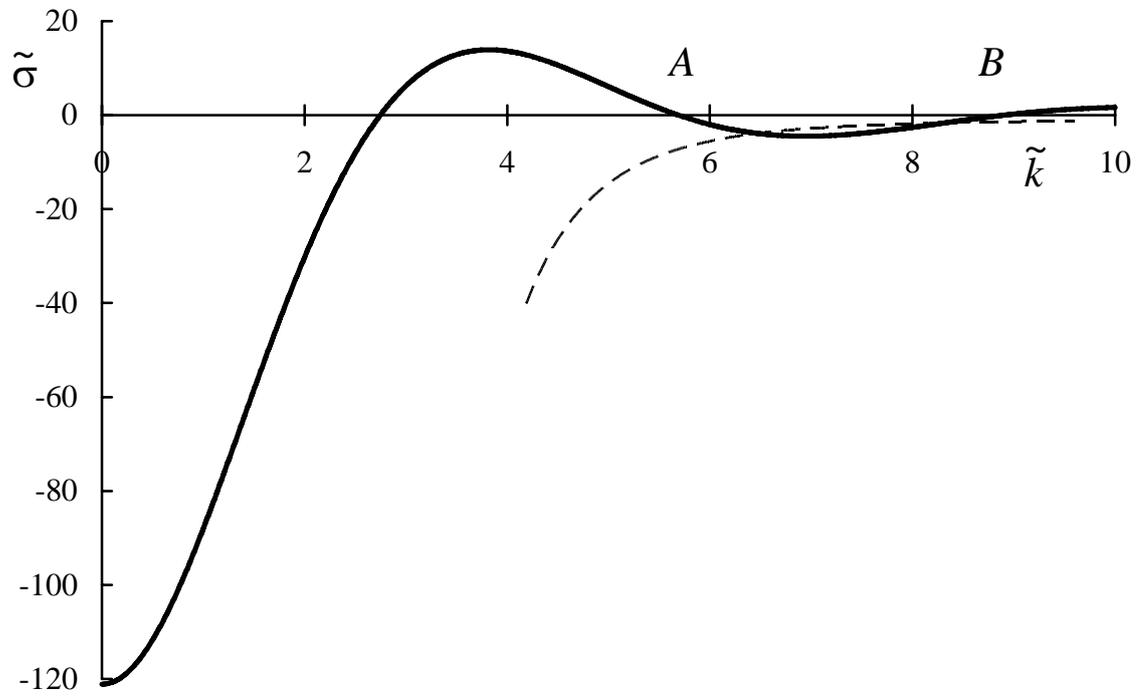

**Figure 2**. Function σ(*k*); the broken curve is explained in Sec. 4